\documentstyle[12pt]{article}

\def\journal{\topmargin .3in    \oddsidemargin .5in
        \headheight 0pt \headsep 0pt
        \textwidth 5.625in 
        \textheight 8.25in 
        \marginparwidth 1.5in
        \parindent 2em
        \parskip .5ex plus .1ex         \jot = 1.5ex}

\journal

\catcode`\@=11
\def\marginnote#1{}

\def\call{{\cal L}}

\def\ie{{\it i.e.}}

\def\be{\begin{equation}}
\def\ee{\end{equation}}
\def\bea{\begin{eqnarray}}
\def\eea{\end{eqnarray}}

\begin{document}
\begin{titlepage}
\begin{center}
\hfill CERN--TH/95--44
\end{center}
\vskip .03in
\center{\Large\bf Non-Unified Sparticle and Particle Masses}
\center{\Large\bf in Unified Theories}
 \vskip .6in
\center{{\sc Savas Dimopoulos~\footnote{On leave of absence
from the Physics Department, Stanford University, Stanford CA 94305,
USA.} and Alex
Pomarol}}
\center{{\it Theory Division, CERN}\\{\it CH-1211 Geneva 23,
Switzerland}}
\vskip .6in
\begin{abstract}
We give examples of minimal extensions of the simplest SU(5) SUSY-GUT
in
which  all squarks and sleptons of a family have different tree level
masses at the
unification scale. This phenomenon is general; it occurs when the
quarks and leptons are the light remnants of a theory which contains
extra heavy families at the unification scale.
The examples have interesting relations between
Yukawa couplings: In one model the
ratio
of the top to bottom Yukawas is as large as $\simeq 3$, partly
accounting for  the large $m_t /m_b$.
Another gives $m_b/m_\tau$ between 2/3 and 1; this relaxes the strict
bounds
on the top mass and  neutrino properties that come from $b$--$\tau$
unification.
Still another allows $ m_s/m_\mu$ to be between 1/6 and 1
and evades the potentially problematic GUT relation of $m_s=m_\mu$.
The final example has horizontal sparticle splittings in spite of the
existence of horizontal symmetries.
\vskip .3in
\noindent CERN--TH/95--44\hfill\\
\noindent February 1995\hfill
\end{abstract}
\end{titlepage}

\section{Sparticle masses as a probe of unification}

It is unlikely that we will ever build a microscope that allows us to
directly study physics at Planckean distances $\sim 10^{-33}$ cm.
The
best we hope for is to directly look at physics near the TeV scale
and
then use theoretical ideas, such as the desert hypothesis, to
extrapolate further. The desert hypothesis acts as a microscope which
magnifies by 14 orders of magnitude without running into any
obstacles
and allows us to look at Planckean distances. Conversely, the
desert hypothesis allows theoretical ideas near the Planck mass to be
translated into low-energy predictions that can be directly compared
to
experiment.
This speculative program has had a remarkable  success: it  predicted
a value for the weak mixing angle in
supersymmetric unified theories (SUSY-GUTs) \cite{dg,drw1,sin2} which
was
subsequently confirmed by experiment \cite{lp}. This gave a
vote of confidence to the desert hypothesis, the existence of light
sparticles and the idea of gauge coupling unification. Of course,
there
are many more parameters in addition to gauge couplings: fermion and
sparticle masses and mixing angles add up to a grand total of 110
physical
parameters
just in the supersymmetric flavor sector. According to the desert
hypothesis, each of  these carries direct information about the
structure of the theory at Planckean distances.

One  objective of this paper is to present some counterexamples to
simple expectations for the
sparticle masses in large classes of unified theories. The
earliest hypothesis on sparticle masses is called universality; it
was
motivated from the need to suppress flavor violating processes
\cite{dg}
 and postulates that all squarks and sleptons
 are degenerate at the unification
scale. Although this strong form of universality is neither
necessary nor likely, it is widely believed that a more restricted
form
is always valid in unified theories:  sparticles belonging to the
same
SU(5) multiplet are
degenerate at the unification scale. This is considered a direct
consequence of unification which will be experimentally checked if
and
when sparticles are discovered at LHC and NLC and their masses are
known with some precision. Furthermore, since the  15 sparticle
species
of 3 families fall into 6 SU(5)-multiplets we should have 9
non-trivial
relations that test unification by looking at sparticle masses,
according to this belief.

We demonstrate that, because the GUT group is spontaneously broken,
there is no good reason for this belief. Sparticles, such
as $\tilde b_R$ and $\tilde\tau_L$,  which by virtue
of
their gauge and family quantum numbers can be grouped into an
irreducible SU(5) multiplet do not necessarily originate from {\it one}
such multiplet; they could come about from a linear combination of
several identical multiplets. This can produce non-degeneracy among
sparticles of the same generation and complementary SU(5) quantum
numbers \footnote{These are sparticles whose combined quantum numbers
complete an irreducible SU(5) multiplet.}. The examples that we
present are not just of mathematical interest; the sparticle
splittings occur precisely in theories (and for the same reasons)
that produce interesting and
desirable relations among fermion masses.

An essential ingredient for these effects is the non-universality of
sparticle masses belonging to {\it different} SU(5) multiplets. This
is
so generic that it hardly needs to be justified. Sparticles in
different
multiplets of the unified group
 have no symmetry reason to be degenerate at the
unification scale or any other scale; even if they are assumed to be
degenerate at the Planck scale their different interactions will
split
them by the time they reach the unification mass \cite{hkr,pp}.
These
splittings are typically large due to the large size of the
representations in unified groups \cite{pp}
\footnote{Such splittings also occur in string
 theories \cite{string}.}. Intermultiplet sparticle splittings can
induce intramultiplet splittings in  higher rank gauge groups such as
SO(10), E(6)  and beyond \cite{dterms}; this mechanism, in contrast to
the one presented here, is not operative in SU(5).

\section{A model with  ${\bf 2/3<m_b/m_\tau<1}$}

The model is a minimal extension of the SU(5) SUSY-GUT
\cite{dg}. Consider a SU(5) theory with just the
 third generation \footnote{Adding the two
 light generations will not change the
conclusions.} consisting of a $\bar{\bf 5}_1$ and ${\bf 10}_1$, the
usual
Higgs fiveplet ${\bf H}$ and antifiveplet $\overline{\bf H}$,
 and the adjoint
${\bf
24}$ that breaks SU(5) down to SU(3)$\times$SU(2)$\times$U(1) at the
unification scale $M_G$  by acquiring a vacuum expectation
value (VEV) that points in the
hypercharge direction:
\be
\langle {\bf 24}\rangle =V_{24}{\bf Y}\equiv V_{24}\, {\rm diag}
(2,2,2,-3,-3)\, .
\label{vevsigma}
\ee
The bottom and tau masses are given by the superpotential
\be
W=h\,
{\bf 10}_1 \overline{\bf H}\,  \bar{\bf 5}_1\, ,
\ee
and are equal at the unification scale \cite{ceg}.
Now add an extra fiveplet and antifiveplet denoted by ${\bf 5}_H$
and
$\bar{\bf 5}_2$    with the following couplings:
\be
W={\bf 5}_H\left[M\,\bar{\bf  5}_1+\lambda\, {\bf 24}\, \bar{\bf
5}_2\right]
+h\, {\bf 10}_1 \overline{\bf H}\,  \bar{\bf 5}_1\, ,
\label{superpotential}
\ee
where $M$ is near the unification mass. One linear combination of
 $\bar{\bf 5}_1$ and
 $\bar{\bf 5}_2$ will acquire a large mass of order $\sim
M_{G}$. The orthogonal combination will be part of the low-energy
spectrum. It contains the right-handed bottom quark and the tau
lepton
doublet which are denoted by $D^c$ and $L$ respectively;  because the
hypercharges of $D^c$ and $L$
differ, it follows from eqs.~(\ref{vevsigma})
and (\ref{superpotential}) that they will be {\it different}
linear combinations of the corresponding states in  $\bar{\bf 5}_1$
and
$\bar{\bf 5}_2$:
\be
\left(\matrix{D^c\cr L}\right)=-\sin\theta_Y \bar{\bf 5}_1
+\cos\theta_Y \bar{\bf  5}_2\, ,
\label{rotation}
\ee
where
\be
\sin\theta_Y=\frac{\rho{\bf Y}}{\sqrt{1+\rho^2 {\bf Y}^2}}\, ,
\ee
with $\rho=\lambda V_{24}/M$.

Since  $\bar{\bf 5}_1$ and $\bar{\bf 5}_2$ are in different
representations
of  SU(5), they have, in general, different soft SUSY breaking masses
at $M_G$ \footnote{Even  if they were equal near the Planck scale,
renormalization effects can induce large splittings at $M_G$
\cite{hkr,pp}.
}:
\be
\call_{soft}=m^2_1|\bar{\bf 5}_1|^2+m^2_2|\bar{\bf 5}_2|^2\, .
\ee
Since the  light combination is given by (\ref{rotation}), one has
\bea
m^2_{\tilde b_R}&=&m^2_2+ s^2_{b_R}(m^2_1-m^2_2)\, ,\nonumber\\
m^2_{\tilde\tau_L}&=&m^2_2+ s^2_{\tau_L}(m^2_1-m^2_2)\, ,
\label{splitst}
\eea
where $b_R\in D^c$, $\tau_L\in L$ and $s_a$  is given by
\be
s_a=\sin\theta_{Y_a}=\frac{\rho{ Y_a}}{\sqrt{1+\rho^2 {Y_a^2}}}\, ,
\ee
and $Y_a$ is the hypercharge of $a$.
Therefore, the squark and slepton masses differ at $M_G$;
their fractional mass-splitting, for $\Delta>0$, is given by
\be
\frac{m^2_{\tilde\tau_L}-m^2_{\tilde b_R}}{m^2_{\tilde\tau_L}}
=\frac{s^2_{\tau_L}-s^2_{b_R}}{\Delta+s^2_{\tau_L}}\,
,
\label{splits}
\ee
where $\Delta=m^2_2/(m^2_1-m^2_2)$.
Eq.~(\ref{splits}) is plotted in Fig.~1 as a function of $\rho$ and
for different values of $\Delta$. It increases for
 small values of $\Delta$. For $\Delta\sim 0.2$
 and $\rho\sim 0.3$, the fractional mass-splitting
is  $\sim 30\%$.

The fermion masses arise from  the Yukawa  coupling
${\bf 10}_1\overline{\bf H}\,\bar{\bf 5}_1$:
\bea
m_b&=&hs_{b_R}\langle\overline{\bf H}\rangle\, ,\nonumber\\
m_\tau&=&hs_{\tau_L}\langle\overline{\bf H}\rangle\, ,
\eea
that leads to the  ratio between the bottom and
tau mass
\be
\frac{m_b}{m_\tau}=\frac{s_{b_R}}{s_{\tau_L}}=
\frac{2}{3}\sqrt{\frac{1+9\rho^2}{1+4\rho^2}}\, .
\label{splitf}
\ee
This ratio  tends to $2/3$ and $1$
in the small and large  $\rho$ limit respectively.
{}From eqs.~(\ref{splits}) and (\ref{splitf}), one can see that the
scalar  mass-splitting is correlated to the fermion mass-splitting.
This is shown in Fig.~1 where the dashed line represents the ratio
$m_b/m_\tau$. The maximum values for the scalar mass-splitting
correspond to
 $m_b/m_\tau\sim 0.7-0.8$.

Thus this model, although it is a minimal perturbation of the SU(5)
SUSY-GUT \cite{dg}, easily accommodates values of $m_b/m_\tau$
 that are between $2/3$ and
$1$. This has at least two interesting implications:
\begin{itemize}
\item It relaxes the strong constraints on the top mass that arise
from
bottom-tau unification \cite{langacker} and  thus reduces  the
 degree of fine-tuning  that is usually required
in these models
to achieve electroweak symmetry breaking
(because the $\mu$ parameter
is typically larger than $m_Z$ \cite{all}).
\item It relaxes the constraints on the neutrino properties that
arise from bottom-tau unification for $\tan\beta <10$
\cite{brignole}.
\end{itemize}

It is now easy to see how the sparticle and particle splittings came
about in this model. Although the right-handed bottom and the tau
lepton doublet -- by virtue of their family and gauge quantum numbers
-- appear to belong to the same $\bar{\bf 5}$ of SU(5), they in fact,
because of their different hypercharges,  came from two {\it
different } linear combinations of a pair of $\bar{\bf 5}$s. This
causes
SU(5) and $SU(4)_{PS}$ breaking effects in sparticles and particles
to be felt at the tree level, since they occur at the very basic
stage of defining the light states of the theory. This mechanism has
been used to produce non-trivial fermion mass
relations for the two light generations \cite{83,adhrs}, but not for
the third family.

\section{A model with  ${\bf 1/6<m_s/m_\mu<1}$}

The main ingredient of the previous section is that the light  states
$D^c$ and $L$ emerge from different linear combinations of states in
two
different $\bar{\bf  5}$s.
The same idea can be implemented for the
states in the decouplets. Add to the previous model an extra
 $\overline{\bf 10}_H$ and  ${\bf 10}_2 $
and, instead of eq.~(\ref{superpotential}),  consider  the
superpotential:
\be
W=\overline{\bf 10}_H\left[M'\,
{{\bf 10}}_1+\lambda'\, {\bf 24\, {\bf 10}}_2\right]+h\,
{\bf 10}_1 \overline{\bf H}\,  \bar{\bf 5}_1\, .
\label{superpotentialb}
\ee
Now the soft masses of the squarks and sleptons embedded in the ${\bf
10}$s
will split  as in eq.~(\ref{splitst})
depending on their hypercharge. As shown in Fig.~2 for
$\Delta m^2/m^2\equiv
(m^2_{\tilde t_R}-m^2_{\tilde t_L})/m^2_{\tilde t_R}$, these
splittings
can be much larger than for the fiveplet since
the differences in hypercharges are larger in the decouplet.

In the fermion sector eq.~(\ref{splitf}) will be modified to
\be
\frac{m_b}{m_\tau}=\frac{s_{b_L}}{s_{\tau_R}}=
\frac{1}{6}\sqrt{\frac{1+36\rho^{'2}}
{1+\rho^{'2}}}\, ,
\label{splitfff}
\ee
where $\rho'=\lambda'V_{24}/M'$. Now the ratio of $m_b/m_\tau$ can
range
from $1/6$ to $1$ and gives even more flexibility to this prediction.
This idea can also be usefully applied to the second generation ratio
$ m_s/m_\mu$ and avoid the potentially problematic
 GUT relation $m_s=m_\mu$.

It is easy to see that if both the fiveplets and the decouplets are
linear combinations of representations, \ie,
the superpotential is the sum of eqs.~(\ref{superpotential})
and (\ref{superpotentialb}), then
 we get the fermion mass ratio
\be
\frac{m_b}{m_\tau}=\frac{s_{b_R}s_{b_L}}{s_{\tau_R}s_{\tau_L}}=
\frac{1}{9}\sqrt{\frac{(1+9\rho^2)(1+36\rho^{'2})}
{(1+4\rho^2)(1+\rho^{'2})}}\, .
\label{splitff}
\ee
This ranges from $1/9$ to 1 and can give even larger particle
mass-splittings.

\section{A SO(10) model with ${\bf 1<h_t/h_b<3}$}

The previous  examples  can be easily adapted to SO(10).
The Higgs sector consists of only two multiplets,
${\bf 45}\ni {\bf 24}$ and  ${\bf 10}\ni \{{\bf H},\, \overline{\bf
H}\}$,
and the matter fields are embedded in the ${\bf 16}$s spinor
representations.
Since the ${\bf 45}$ is in the adjoint representation of SO(10),
its VEV can point in any direction of the two dimensional subspace of
SO(10) generators  that commutes
 with SU(3)$\times$SU(2)$\times$U(1).
One  possible direction, as in the previous example, is the
hypercharge
direction ${\bf Y}$. Here we will consider the case where the VEV
of the ${\bf 45}$ points in the   ${\bf X}$-direction, where ${\bf X}$
is the SO(10)
generator
that commutes with the SU(5) subgroup, \ie
 $\langle {\bf 45}\rangle=V_{45}{\bf X} $ \footnote{Since this
direction preserves SU(5), we still
have bottom-tau unification.}.
The superpotential in now given by
\be
W=\overline{\bf 16}_H\left[M {\bf 16}_1+\lambda\, {\bf 45\,
16}_2\right]+h\,
{\bf 16}_2{\bf 10\,  16}_2\, .
\label{superpotentialc}
\ee
Notice we have now coupled the ${\bf 16}_2$ instead of the ${\bf
16}_1$
to the Higgs ${\bf 10}$.
As in the previous model, the light quarks and leptons
arise from  the linear combination,
 $-\sin\theta_X{\bf 16}_1+\cos\theta_X{\bf 16}_2$, where
now the mixing angle is
\be
\sin\theta_X=\frac{\rho{\bf X}}{\sqrt{1+\rho^2 {\bf X}^2}}\, ,
\ee
where $\rho=\lambda V_{45}/M$. The
$X$-charges of the quarks and leptons are
\be
 X_a=(1,1,-3,-3,1,5)\ \ {\rm for\ \ }a=(Q,U^c,D^c,L,E^c,\nu^c)\, ,
\label{xnumbers}
\ee
where $Q$  is  the quark  SU(2)-doublet
and $U^c$, $E^c$ and $\nu^c$ are  the quark and lepton singlets.
The scalar masses are split according to
\be
m^2_{\tilde a}=m^2_2+ s^2_{a}(m^2_1-m^2_2)\, ,
\label{splitsb}
\ee
where $s_a\equiv\sin\theta_{X_a}$ and $m^2_i$ is
 the soft mass of ${\bf 16}_i$.
These scalar mass-splittings
 preserve SU(5)
  since the $\langle {\bf 45}\rangle$ does not break this subgroup of
SO(10).

The fermion masses in this model are proportional to
$c_a=\sqrt{1-s^2_a}$
since
 we coupled  the Higgs ${\bf 10}$
to the  ${\bf 16}_2$.
For the third family, we have
\be
\frac{h_t}{h_b}=\frac{c_{t_R}c_{t_L}}{c_{b_R}c_{b_L}}=
\sqrt{\frac{1+9\rho^2}{1+\rho^2}}\, ,
\label{splitfb}
\ee
which has a maximum value of $3$ for $\rho\rightarrow\infty$.
Nevertheless,
we cannot  take very large values of $\rho$
since the top mass is given by
\be
m_t= \frac{h\langle {\bf 10}\rangle}{1+\rho^2}\, ,
\ee
and decreases  when increasing $\rho$.
To avoid a light top or a too large $h$-Yukawa coupling,
the value of  $\rho$ cannot be larger than $1-2$ that leads to
 $h_t/h_b\sim 2.2-2.7$.

Therefore this minimal extension of the simplest SO(10) theory can
lead to a partial explanation of the large top-bottom mass ratio,
allows for moderate values of $\tan\beta \sim 20 $ and (since
$h_t>h_b$)
 accommodates radiative electroweak breaking without an extreme
fine-tuning. In contrast the minimal SO(10) theory  has
$h_t=h_b=h_\tau=h_\nu$; this implies that  $\tan\beta$ is large,
$\tan\beta\sim 40-60$, and a severe fine-tuning is required to get the
correct
electroweak symmetry breaking \cite{hrs}. Another difference with the
minimal
SO(10) theory  has to do with neutrinos. In our theory,  from
eq.~(\ref{xnumbers}), one has that the Dirac-type Yukawa coupling of
the neutrino  is very suppressed
\be
\frac{h_\nu}{h_t}=\frac{c_{\nu_R}c_{\nu_L}}{c_{t_R}c_{t_L}}
=\frac{1+\rho^2}{\sqrt{(1+25\rho^2)(1+9\rho^2)}}\, ,
\label{splitfc}
\ee
and  therefore does not modify  the value of $m_b/m_\tau$  even if
the right-handed neutrino mass is light $\ll 10^{16}$ GeV and
$ \nu_L$ is cosmologically interesting.  In contrast in  SO(10) with
$h_t=h_\nu$, the neutrino coupling affects  $m_b/m_\tau$; this
 effectively
excludes neutrinos  from being cosmologically relevant for values of
$\tan\beta<10$ \cite{brignole}. Finally, to accommodate
$m_b/m_\tau$  our theory  disfavors large values for the strong
coupling
and  small values for the bottom mass --unless there are
significant low-energy contributions to $m_b$ \cite{hrs,carena}.

\section{Horizontal splittings with horizontal symmetries}

The  mechanism of this paper
can also generate  horizontal mass-splittings in
 theories with  horizontal symmetries.
As an example, consider a
SO(10)$\times$SU(3)$_H$ theory where the SU(3)$_H$ is a horizontal
symmetry under which the three families
 transform as a triplet.
The superpotential is given by
\bea
W&=&(\overline{\bf16},\bar{\bf 3})_H\left[\, M\, ({\bf 16},{\bf 3})_1
+\lambda\, ({\bf 45},{\bf 8})\,  ({\bf 16},{\bf
3})_2\,\right]\nonumber\\
&+&h\,({\bf 16},{\bf 3})_2({\bf 10},\bar{\bf 6})\,  ({\bf 16},{\bf
3})_2\, .
\label{superpotentiald}
\eea
Depending on the direction of the
VEV of the $({\bf 45},{\bf 8})$, we have different possibilities of
splitting the masses of the three families.
Constraints from flavor violating processes
 requires near degeneracy between the first and second family
scalars \cite{dg};
this can be guaranteed if the
VEV of the $({\bf 45},{\bf 8})$ points in the direction
 $T_8={\rm diag}(1,1,-2)$ of
 SU(3)$_H$ which preserves
SU(2)$_H$.
If  respect SO(10), the
VEV of  $({\bf 45},{\bf 8})$
 points in the ${\bf X}$-direction \footnote{The fermion mass
relations in this model will be like those of the previous section.
Since these relations  work  best for the third family,
it is simplest to assume that
only $({\bf 10},\bar{\bf 6}_{33})$
is light.},
 the sparticle masses are given by eq.~(\ref{splitsb})
where now the mixing angle is
given by
\be
s_{a(i)}=\frac{\rho  X_a[T_8]_i}{\sqrt{1+\rho^2
{X_a^2[T_8]^2_i}}}\, ,
\ee
where $X_a$ is given in eq.~(\ref{xnumbers})
and $[T_8]_i=(1,1,-2)$ for $i=(1^{\rm st},2^{\rm nd},3^{\rm rd}\, {\rm
family})$.
The sparticles of the third family are split
from those of the first and second families because the SU(3)$_H$
symmetry is broken.

\section{Summary}

We presented a mechanism that leads to interesting splittings of
particle and sparticle masses at the same time. It implies that
sparticles belonging to the {\it same} generation can have significant
splittings at the unification scale.  How general is the
mechanism we presented? It occurs in a large class of theories which,
near the
unification mass,
have $N+3$ left-handed and $N$ right-handed families  which
subsequently
combine to leave 3 light families. Because the
physics that determines the light states that we call quarks
and leptons breaks the unified group, the tree
level particle and sparticle masses in general do not obey any GUT
relations \footnote{ If
they do, it is an indication of some unexpected
simplicity that should be understood.  The absence of
large flavor violations is an indication of partial
simplicity in the horizontal direction for some of the sparticles of
the two
lightest generations.} and lead to splittings. Therefore the sparticle
spectroscopy that results is richer and may provide us with more
detailed information about the physics near the unification mass than
in minimal theories with universality.
\vskip .25in
\noindent {\large{\bf Acknowledgements}}
\vskip .1in
We thank M. Carena and C. Wagner for discussions
 on $b$--$\tau$ unification.

\vskip .5in

\noindent{\Large{\bf Figure Captions}}
\vskip .2in
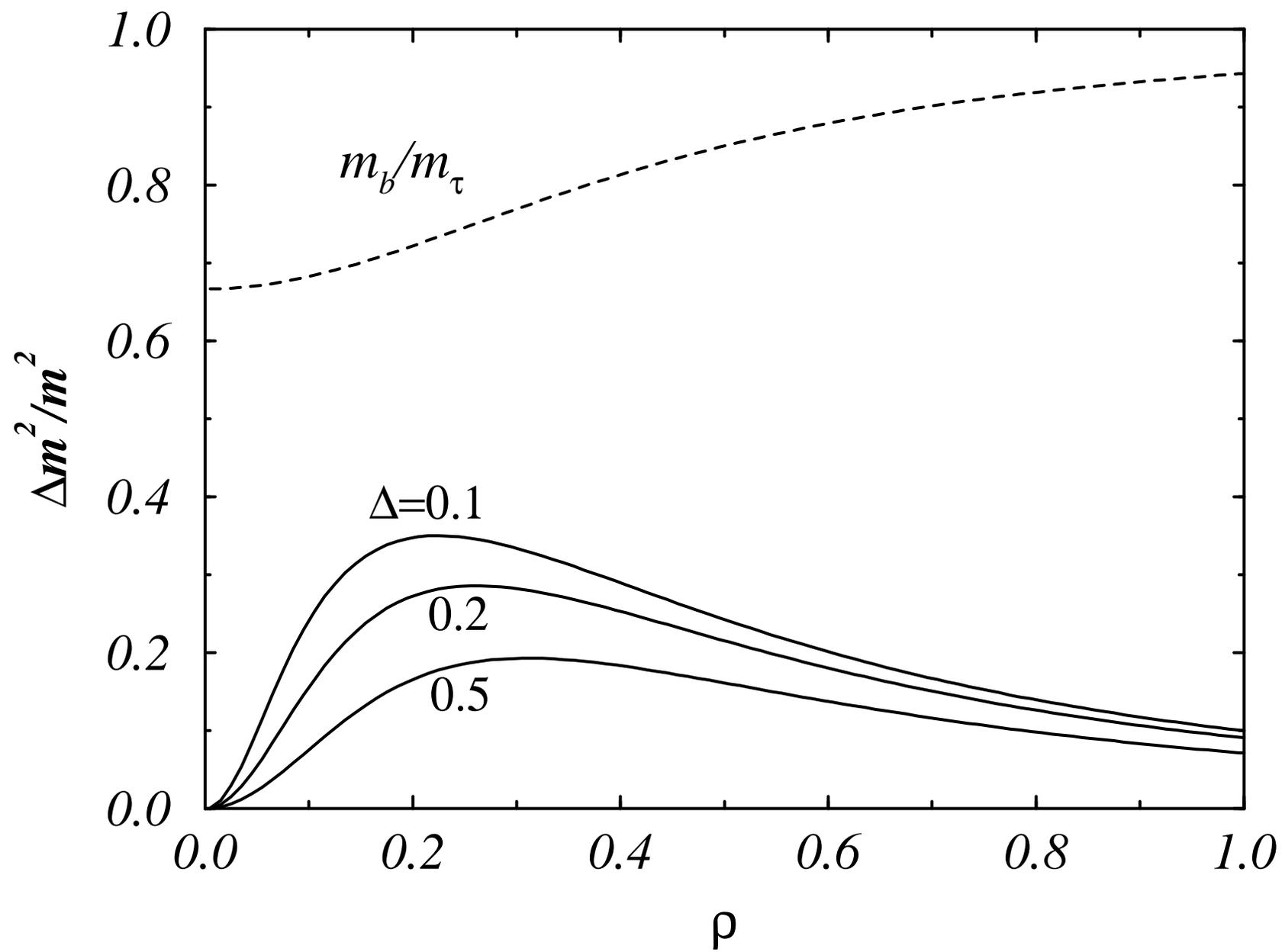
\noindent{\bf Fig.~1:} {Scalar mass-splitting $\Delta m^2/m^2\equiv
(m^2_{\tilde\tau_L}-m^2_{\tilde b_R})/m^2_{\tilde\tau_L}$
 as a function of $\rho=\lambda V_{24}/M$ and for
different values of $\Delta=m^2_2/(m^2_1-m^2_2)$. The dashed line
corresponds to the the ratio $m_b/m_\tau$.}
\vskip .07in
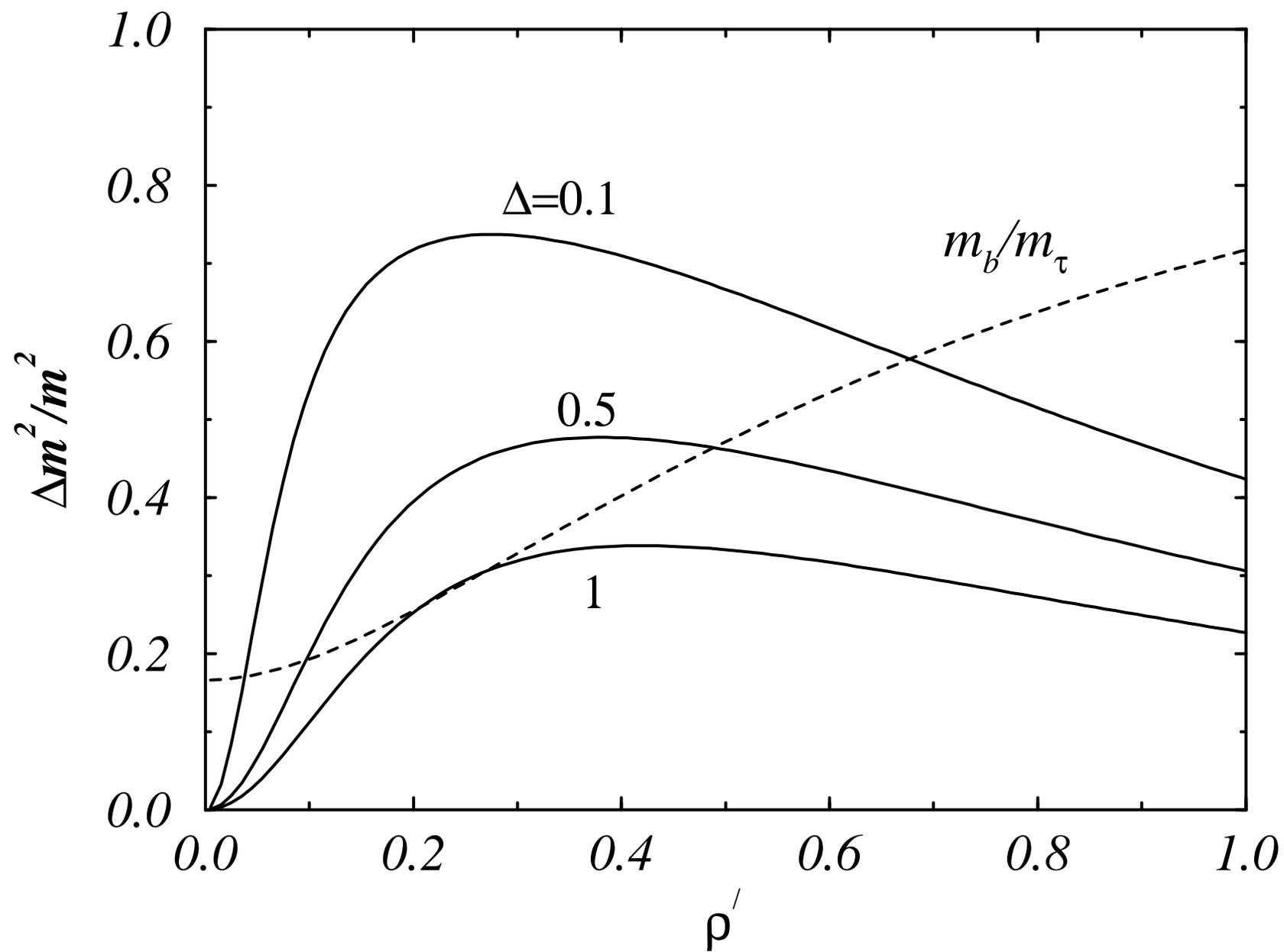
\noindent{\bf Fig.~2:} {Scalar mass-splitting $\Delta m^2/m^2\equiv
(m^2_{\tilde t_R}-m^2_{\tilde t_L})/m^2_{\tilde t_R}$
 as a function of $\rho'=\lambda' V_{24}/M'$ and for
different values of $\Delta=m^2_2/(m^2_1-m^2_2)$. The dashed line
corresponds to the the ratio $m_b/m_\tau$.}
\pagebreak

\begin{thebibliography}{99}
\bibitem{dg} S. Dimopoulos and H. Georgi, ``Supersymmetric
GUTs'', p. 285, {\it Second Workshop on Grand Unification}, University
of
Michigan, Ann Arbor, April 24--26 (1981), eds. J. Leveille, L. Sulak
and
D. Unger, Birkhauser (1981);
S. Dimopoulos and H. Georgi, Nucl. Phys. {\bf B193} (1981) 150.
\bibitem{drw1} S. Dimopoulos, S. Raby and F. Wilczek, Phys.
Rev. {\bf D24} (1981) 1681.
\bibitem{sin2} N. Sakai, Z. Phys. {\bf C11} (1981) 153; L. Iba\~nez and
G.G. Ross, Phys. Lett. {\bf B105} (1981) 439; M.B. Einhorn and
D.R.T. Jones, Nucl. Phys. {\bf B196} (1982) 475; W.J. Marciano
and G. Senjanovic, Phys. Rev. {\bf D25} (1982) 3092.
\bibitem{lp}Excellent recent analyses are P. Langacker and N.
Polonsky, Phys. Rev. {\bf D47} (1993) 4028;
 L.J. Hall and U. Sarid, Phys. Rev. Lett. {\bf 70} (1993) 2673.
\bibitem{hkr}
L.J. Hall, V.A. Kostelecky and S. Raby, Nucl. Phys. {\bf B267} (1986)
415;\\
H. Georgi, Phys. Lett. {\bf B169} (1986) 231.
\bibitem{pp}
N. Polonsky and A. Pomarol, Phys. Rev. Lett. {\bf 73} (1994) 2292;
Preprint hep--ph/9410231.
\bibitem{string}
L. Iba\~nez and D. Lust, Nucl. Phys. {\bf B382} (1992) 305;
V. Kaplunovsky and J. Louis, Phys. Lett.  {\bf B306} (1993) 269;
A. Brignole, L. Iba\~nez and C. Mu\~noz,
Nucl. Phys. {\bf B422} (1994) 125;
S. Ferrara, C. Kounnas and F. Zwirner, CERN Preprint TH--7192--94;
 J. Louis and Y. Nir, M\"unchen Preprint LMU--TPW 94--17.
\bibitem{dterms}
M. Drees, Phys. Lett. {\bf B181} (1986) 279;
J. S. Hagelin and S. Kelley, Nucl. Phys. {\bf B342} (1990) 95;
Y. Kawamura, H. Murayama and M. Yamaguchi,
Phys. Lett.  {\bf B324} (1994) 52; H.C. Cheng and L.J. Hall,
Berkeley Preprint LBL--35950.
\bibitem{ceg}
M.S. Chanowitz, J. Ellis and M.K. Gaillard,
 Nucl. Phys. {\bf B128} (1977) 506.
\bibitem{langacker}
See for example,
P. Langacker and N. Polonsky, Phys. Rev. {\bf D49} (1994) 1454.
\bibitem{all}
See for example,
V. Barger, M.S. Berger and P. Ohmann, Phys. Rev. {\bf D49} (1994) 4908;
M. Carena, M. Olechowski, S. Pokorski and C.E.M. Wagner,
Nucl. Phys. {\bf B419} (1994) 213; and Ref.~\cite{pp}.
\bibitem{brignole}
F. Vissani and A.Y. Smirnov, SISSA Preprint SISSA--63/94/EP;
A. Brignole, H. Murayama and R. Rattazzi, Berkeley Preprint
LBL--35774.
\bibitem{83}
S. Dimopoulos, Phys. Lett. {\bf B129} (1983) 417.
\bibitem{adhrs}
G. Anderson, S. Dimopoulos, L.J. Hall, S. Raby and G.D. Starkman,
 Phys. Rev. {\bf D49} (1994) 3660.
\bibitem{hrs}
L.J. Hall, R. Rattazzi and U. Sarid,  Phys. Rev. {\bf D50} (1994) 7048;
Preprint hep--ph/9405313.
\bibitem{carena}
M. Carena, M. Olechowski, S. Pokorski and C.E.M. Wagner,
Nucl. Phys. {\bf B426} (1994) 269.
\end{thebibliography}
\end{document}